\documentclass[9pt,conference]{IEEEtran}
\IEEEoverridecommandlockouts

\usepackage{cite}

\usepackage{amsmath,amssymb,amsfonts}
\usepackage{graphicx}
\usepackage{textcomp}
\usepackage{xcolor}
\usepackage{booktabs}
\usepackage{microtype}
\usepackage{tikz}
\usepackage{ulem}
\usetikzlibrary{shapes.geometric}
\usepackage{tikz-cd}
 \usetikzlibrary{calc}
\def\BibTeX{{\rm B\kern-.05em{\sc i\kern-.025em b}\kern-.08em
    T\kern-.1667em\lower.7ex\hbox{E}\kern-.125emX}}

\usepackage{titlesec}
\titlespacing{\section}{0pt}{6pt}{3pt}
\titlespacing{\subsection}{0pt}{3.5pt}{2.5pt}

\usepackage[nolist]{acronym}

\begin{acronym}
\acro{stft}[STFT]{short-time Fourier transform}
\acro{istft}[iSTFT]{inverse short-time Fourier transform}
\acro{dnn}[DNN]{deep neural network}
\acro{pesq}[PESQ]{Perceptual Evaluation of Speech Quality}
\acro{polqa}[POLQA]{perceptual objectve listening quality analysis}
\acro{wpe}[WPE]{weighted prediction error}
\acro{psd}[PSD]{power spectral density}
\acro{rir}[RIR]{room impulse response}
\acro{hrir}[HRIR]{head-related impulse response}
\acro{brir}[BRIR]{binaural room impulse response}
\acro{hrtf}[HRTF]{head-related transfer function}
\acro{snr}[SNR]{signal-to-noise ratio}
\acro{lstm}[LSTM]{long short-term memory}
\acro{polqa}[POLQA]{Perceptual Objective Listening Quality Analysis}
\acro{sdr}[SDR]{signal-to-distortion ratio}
\acro{cnns}[CNNs]{Convolutional Neural Networks}
\acro{vae}[VAE]{variational auto-encoder}
\acro{gan}[GAN]{generative adversarial network}
\acro{tf}[T-F]{time-frequency}
\acro{sde}[SDE]{stochastic differential equation}
\acro{ode}[ODE]{ordinary differential equation}
\acro{drr}[DRR]{direct to reverberant ratio}
\acro{lsd}[LSD]{log spectral distance}
\acro{sisdr}[SI-SDR]{scale-invariant signal to distortion ratio}
\acro{mos}[MOS]{mean opinion score}
\acro{map}[MAP]{maximum a posteriori}
\acro{sde}[SDE]{stochastic differential equation}
\acro{ode}[ODE]{ordinary differential equation}
\acro{dps}[DPS]{diffusion posterior sampling}
\acro{tfrs}[TFRs]{Time-Frequency Representations}
\acro{cqt}[C$Q$T]{Constant-$Q$ Transform}
\acro{ldms}[LDMs]{Latent Diffusion Models}
\acro{shd}[SHD]{Spherical Harmonics Decomposition}
\acro{hoa}[HOA]{Higher-Order Ambisonics}
\acro{rir}[RIR]{Room Impulse Response}
\acro{atf}[ATF]{Array Transfer Function}
\acro{ls}[LS]{Least-Squares}
\end{acronym}
\begin{document}

\title{Ambisonics Encoding of Room Impulse Responses using a Device-Agnostic Diffusion Model}

\author{
\begin{tabular}{@{}c@{}}
Eloi Moliner$^{1,2}$
\qquad  Christoph Hold$^{1}$
\qquad Juan Azcarreta Ortiz$^{1}$
\qquad Sebastian Prepeliţă$^{1}$\\
\qquad Ishwarya Ananthabhotla$^{1}$
\qquad Daniel Wong$^{1}$
\qquad Sanjeel Parekh$^{1}$
\qquad Sanha Lee$^{1}$
\end{tabular}
\IEEEauthorblockA{ \\
\\
$1$ Reality Labs Research, Meta \quad $2$ Acoustics Lab, DICE, Aalto University, Espoo, Finland \\
}
}

  

\maketitle

\begin{abstract}
We address the problem of encoding room impulse responses (RIRs) into high-order Ambisonics (HOA) representations from arbitrary and potentially insufficient or incomplete microphone array measurements. This task is fundamentally ill-posed for microphone arrays with limited spatial capture capabilities, such as irregular or sparse arrays, as classical linear methods fail to reconstruct high-order spatial detail.
 We introduce a diffusion-based generative framework that models the statistical properties of HOA RIRs. 
 This enables device-agnostic encoding from arbitrary microphone arrays, potentially unseen during data measurement.
 Our approach incorporates a posterior sampling procedure that enforces consistency between the estimated signals and the measurements while plausibly reconstructing spatial information that is unobservable from the limited measurements alone. Experiments on simulated data demonstrate that our method outperforms linear and neural baselines, achieving accurate HOA RIR estimation up to 12th order. 
 A listening test with binaural renderings, including both simulated and measured RIRs, further confirms that the proposed method yields higher perceptual similarity to reference Ambisonics RIRs than all baselines.
The flexibility and accuracy of the proposed framework opens new possibilities for scalable acoustics simulations.
\end{abstract}


\vspace{-2pt}
\section{Introduction}
\vspace{-2pt}
\label{sec:intro}

Ambisonics provides a compact representation of a spherical sound field at the point of observation via \ac{shd}, enabling flexible rendering to arbitrary listening / head directions, and playback configurations \cite{zotter2019ambisonics}.
When the sound field is represented using higher \ac{shd} orders, the \ac{hoa} signal can express finer spatial detail. However, accurate \ac{hoa} acquisition typically requires dense spherical sampling \cite{rafaely2015fundamentals}.
Estimating \ac{hoa} signals from irregular microphone arrays, such as wearable devices, is challenging because the available measurements often lack sufficient spatial bandwidth to resolve higher-order components \cite{rafaely2015fundamentals}.

This work focuses on Ambisonics encoding of \acp{rir} from microphone-array measurements. \acp{rir} encapsulate the acoustic propagation and are central to downstream spatial audio applications, including acoustic simulation, binaural and loudspeaker rendering, and data generation for training and evaluating speech enhancement and spatial audio models. Representing \acp{rir} in the \ac{hoa} domain is particularly attractive because it disentangles the room acoustics (captured as \ac{shd} coefficients) from the specifics of any given recording device: once an \ac{hoa} \ac{rir} is available, the soundfield can be rotated arbitrarily (the \ac{shd} is directionally continuous) and sampled/decoded at arbitrary directions. Furthermore, since the \ac{shd} is device-agnostic, we may map the \ac{shd} \ac{rir} to microphone signals of different microphone arrays
given their \acp{atf}, by imprinting new acoustic scattering and sampling properties. This ability is valuable for scalable data generation, cross-device evaluation, and rapid iteration over array designs without repeating time-consuming measurement campaigns.

Classical Ambisonics encoding methods typically implement a linear least-squares projection from microphone signals to \ac{shd} coefficients \cite{politis2017comparing}. For irregular microphone arrays, the encoding operator often becomes poorly conditioned or rank-deficient, limiting the spatial detail that can be reconstructed from the available signals. Prior work has introduced additional structure into the inverse problem through model assumptions or learned priors, including signal sparsity \cite{wabnitz2011upscaling} and parametric spatial audio \cite{mccormack2022parametric}. However, these approaches are primarily designed for perceptual relevance and rely on assumptions that may limit reconstruction accuracy when violated.



Data-driven methods based on deep neural networks have recently been proposed for related tasks. Neural Ambisonics encoders predict time-dependent encoding matrices from specific microphone arrays to Ambisonics formats \cite{heikkinen2024neural, deppisch2026residual}, but are typically limited to first-order and device-specific cases, despite efforts to improve robustness across arrays \cite{heikkinen2026beyond}. Ambisonics order upmixing, such as extrapolating First-Order Ambisonics (FOA) to \ac{hoa}, has also been explored using parametric spatial audio methods \cite{politis2017comparing} and also predictive neural networks \cite{nawfal2025ambisonics} and generative diffusion models \cite{milstein2025diffaudiffusionbasedambisonicsupscaling}
Importantly, both neural approaches have focused on scene-based audio signals, with no demonstrated results for accurate \ac{rir} encoding.


In this work, we formulate the Ambisonics encoding of array \acp{rir} as an inverse problem and address it within a Bayesian probabilistic framework. The ill-posed nature of the task motivates the use of generative models, which can sample from learned posterior distributions rather than producing a single deterministic estimate. Diffusion models have recently emerged as a powerful approach \cite{lai2025principles}, serving as expressive priors for solving inverse problems \cite{daras2024survey, moliner2023solving, xu2025arraydpsunsupervisedblindspeech, lee2026solvingroomimpulseresponse}.
We propose a diffusion model for \ac{hoa} \acp{rir} based on a hybrid architecture that operates jointly in the time–frequency and time domains. Building on this model, we introduce an iterative inference procedure based on diffusion posterior sampling that leverages the device-agnostic generative prior to solve the inverse problem for any array with a known \ac{atf}.
Experiments in simulated environments show improved performance over classical least-squares encoding and neural baselines for \acp{rir} up to 12th-order Ambisonics. Finally, we provide evidence of the perceptual relevance of the proposed approach through the results of a formal listening test conducted with both simulated and measured room impulse responses.

\vspace{-2pt}
\section{Problem Definition}
\vspace{-2pt}
\label{sec:problem_definition}

We consider a source--receiver configuration in a reverberant enclosure, and a microphone array with $M$ microphones. Let
$\mathbf{x}\in\mathbb{R}^{M\times T}$ denote the multichannel \ac{rir} measured at the array, where $T$ is the number of time samples.
The room response around the receiver unit sphere can be represented as a continuum of plane waves with a
directional amplitude density.
A compact representation is obtained by expanding this directional density in spherical harmonics up to order $N$, yielding $(N+1)^2$
time-domain HOA coefficient signals. We collect these coefficients over time in $\mathbf{a}\in\mathbb{R}^{(N+1)^2\times T}$, where each row
corresponds to a spherical-harmonic component $\mathbf{a}_{nm}[t]$ with order $n\in\{0,\dots,N\}$ and degree $m\in\{-n,\dots,n\}$ \cite{rafaely2015fundamentals}.

Let $\mathbf{h}\in\mathbb{R}^{M\times (N+1)^2 \times T}$ denote the \ac{atf} in the time domain, which was obtained on a dense spherical grid of directions and then transformed to the SHD. 
In practice, $\mathbf{h}$ can be obtained from measurements or simulations, and depends on the array geometry, device scattering, and
microphone characteristics.


In the frequency domain, the forward model decouples across frequency bins and simplifies to a matrix multiplication. For each frequency bin $k\in\{0,\dots,K-1\}$,
\vspace{-3pt}
\begin{equation}
\label{eq:forward_matrix}
\mathbf{X}[k] = \mathbf{H}[k]\mathbf{A}[k],
\vspace{-3pt}
\end{equation}
where $\mathbf{X}[k]\in\mathbb{C}^{M}$ is the DFT of the measurements $\mathbf{x}$, $\mathbf{H}[k]\in\mathbb{C}^{M\times (N+1)^2}$ is the DFT of the \ac{atf} $\mathbf{h}$, and $\mathbf{A}[k]\in\mathbb{C}^{(N+1)^2}$ is the DFT of the HOA coefficients $\mathbf{a}$.
For notational simplicity, we omit the frequency index $k$ in subsequent expressions; all matrix multiplications are understood to be carried out independently for each frequency bin.
Given the measurements $\mathbf{x}$ and the \ac{atf} $\mathbf{h}$, the goal is to estimate the Ambisonics coefficients $\mathbf{a}$. 



A standard solution for Ambisonics encoding is the least-squares estimate 
$\hat{\mathbf{A}}_{\mathrm{LS}} = \mathbf{H}^\dagger\mathbf{X}$ \cite{politis2017comparing},  
where $\mathbf{H}^\dagger \in\mathbb{C}^{(N+1)^2\times M}$
denotes the Moore--Penrose pseudoinverse of $\mathbf{H}$ (computed per frequency), and is defined as
$\mathbf{H}^\dagger =\mathbf{H}^H\left(\mathbf{H}\mathbf{H}^H\right)^{-1} $.
When the target Ambisonics order $N$ is high, $(N+1)^2$ can greatly exceed the number of microphones $M$, and $\mathbf{H}$ can become ill-conditioned due to limited aperture and spatial aliasing.
In practice, $\mathbf{H}^\dagger$
is replaced by a regularized inverse that improves numerical conditioning by discarding the effect of small singular values, without resolving the underlying ambiguity \cite{hansen1987truncated}.
In this regime, the least-squares solution cannot recover spatial detail not supported by the array.  




 Since the inverse problem is ill-posed, many HOA RIRs $\mathbf{a}$ can explain the same measurements $\mathbf{x}$. We therefore formulate the task as posterior inference and aim to model the conditional distribution $p(\mathbf{a}\mid \mathbf{x},\mathbf{h})$ using a data-driven generative model.
The posterior can be factorized as:
$p(\mathbf{a}\mid \mathbf{x},\mathbf{h}) \;\propto\; p(\mathbf{a})\,p(\mathbf{x}\mid \mathbf{a},\mathbf{h})$.
Because the SHD representation is device independent, this formulation naturally separates two components.
The prior $p(\mathbf{a})$ captures the statistics of plausible HOA RIRs and can be learned using a generative model trained with a dataset of HOA \acp{rir}.
The likelihood term $p(\mathbf{x}\mid \mathbf{a},\mathbf{h})$ is defined by the known forward model in \eqref{eq:forward_matrix}, which describes how the microphone measurements are generated from the HOA coefficients for a given array transfer function.
At inference time, the two terms are combined: the likelihood enforces consistency with the microphone measurements, while the prior supplies plausible spatial detail that cannot be uniquely determined from the array observations.


\vspace{-2pt}
\section{Methods}

\vspace{-2pt}
\subsection{Diffusion Model Formulation}
\vspace{-1pt}
\label{sec:prior}

We model the distribution of time-domain HOA RIRs $p(\mathbf{a})$ using a generative diffusion model \cite{lai2025principles}. While recent works proposed generative approaches for \acp{rir} \cite{ratnarajah2022fast, arellano2025room,lin2026genchoroomimpulseresponse, lluis2025blind, della2025diffusionrir}, no generative model has been proposed for \ac{hoa} \acp{rir} up to date.
 Diffusion models learn a mapping between a prior distribution (typically a Gaussian) and an empirical data distribution, in our case observed through a dataset of \ac{hoa} RIRs. This mapping is parameterized by a time variable $\tau$.  Generation can be achieved by iteratively traversing the mapping from a high-noise regime toward the data distribution.

Each training example $\mathbf{a}\in\mathbb{R}^{(N+1) ^2 \times T}$ consists of a \ac{hoa} \ac{rir}, padded to a fixed $T$ time samples.
 A continuous set of the noisy marginals $p(\mathbf{a}_\tau)$ is constructed by defining an isotropic diffusion process in the data space, where independent Gaussian noise with identical variance is added to every Ambisonics coefficient and temporal sample. Similar to \cite{karras2022elucidating}, we define the perturbation process
$\mathbf{a}_{\tau} = \mathbf{a} + \tau\,\mathbf{z}$,
where
$\mathbf{z}\sim\mathcal{N}(\mathbf{0},\mathbf{I})$.
Intuitively, increasing $\tau$ yields progressively more corrupted versions of $\mathbf{a}$, and for sufficiently large $\tau$ the distribution of $\mathbf{a}_\tau$ approaches an isotropic Gaussian.


Sampling can be formulated by means of the \textit{probability flow ODE}, which transports samples from a Gaussian distribution toward the data distribution:
\vspace{-2pt}
\begin{equation}\label{eq:probabilityflow}
    \mathrm{d}\mathbf{a}_\tau = - \tau \nabla_{\mathbf{a}_\tau} \log p(\mathbf{a}_\tau) \mathrm{d}\tau .
    \vspace{-2pt}
\end{equation}
This allows generating samples by integrating the ODE down to $\tau\to 0$.
The score function $\nabla_{\mathbf{a}_\tau}\log p(\mathbf{a}_\tau)$ is not available in closed form. Nevertheless, it can be approximated.
Assuming a Gaussian perturbation process, the score can be related to the minimum mean-square error (MMSE) denoiser $\mathbb{E}[\mathbf{a}_0 \mid \mathbf{a}_\tau]$ via \cite{lai2025principles}:
\vspace{-2pt}
\begin{equation}
\nabla_{\mathbf{a}_\tau}\log p(\mathbf{a}_\tau)
=
-(\mathbf{a}_\tau-\mathbb{E}[\mathbf{a}_0\mid \mathbf{a}_\tau])/\tau^2.
\vspace{-2pt}
\end{equation}
Diffusion models approximate the MMSE estimate using a neural network, i.e.,
$\mathbb{E}[\mathbf{a}_0 \mid \mathbf{a}_\tau] \approx \hat{\mathbf{a}}_0(\mathbf{a}_\tau)$.
Similar to Karras et al.~\cite{karras2022elucidating}, we parameterize the denoised prediction with a residual model 
\vspace{-2pt}
\begin{equation}
\label{eq:edm_den}
\hat{\mathbf{a}}_0(\mathbf{a}_\tau)
=
c_{\mathrm{skip}}(\tau)\,\mathbf{a}_\tau
+
c_{\mathrm{out}}(\tau)\,
F_\psi\!\left(
c_{\mathrm{in}}(\tau)\,\mathbf{a}_\tau,\,
\tau
\right),
\vspace{-2pt}
\end{equation}
where $F_\psi$ is a deep neural network, and \(c_{\mathrm{skip}}(\tau)\), \(c_{\mathrm{out}}(\tau)\), \(c_{\mathrm{in}}(\tau)\), and \(c_{\mathrm{noise}}(\tau)\) are scalar weighting functions of the noise level \(\tau\).

Different to \cite{karras2022elucidating}, we train $F_\psi$ using conditional flow matching \cite{lipman2022flow}, which empirically yields  faster optimization in our setting. Specifically, for samples $\mathbf{a}\sim p(\mathbf{a})$, noise $\mathbf{z}\sim\mathcal{N}(\mathbf{0},\mathbf{I})$, and $\tau$ drawn from a training distribution, we minimize
\vspace{-2pt}
\begin{equation}
\label{eq:cfm_obj}
\mathbb{E}_{\mathbf{a},\mathbf{z},\tau}\Big[
\big\|
F_\psi\!\big(c_{\mathrm{in}}(\tau)\,(\mathbf{a}+\tau\mathbf{z}),\,\tau\big)
-
\big(\mathbf{z}-\mathbf{a}\big)
\big\|_2^2
\Big],
\vspace{-2pt}
\end{equation}
This design choice requires setting the preconditioning parameters as 
$c_{\mathrm{in}}(\tau)= c_{\mathrm{skip}}(\tau) = \frac{1}{\tau+1}$, 
$c_{\mathrm{out}}(\tau)  =
\frac{\tau}{\tau+1}$ \cite{schusterbauer2025diff2flow}.

\vspace{-1pt}
\subsection{Hybrid Two-Stage Architecture for HOA RIRs} \label{sec:architecture}
\vspace{-1pt}

An important design choice is the architecture of the backbone model $F_\psi$. In our setting, the model operates on raw Ambisonic RIRs represented with all coefficients stacked along the channel axis. Because this representation is high-dimensional and considering that large-scale datasets are not available, incorporating inductive biases derived from domain knowledge can be advantageous.

Our design is motivated by the distinct structure of RIRs across time. The early part of a RIR is dominated by the direct path and a set of early reflections. Accurately modeling these components is essential to preserve spatial cues, since their relative timing and directional content largely determines the perceived localization. In contrast, the late reverberation is more diffuse and exhibits smoother temporal structure, with frequency-dependent energy decay rates that are naturally captured in the time-frequency domain. 

We adopt a hybrid two-stage architecture that combines time-frequency and time-domain processing, summarized in Fig.~\ref{fig:architecture}.
The first stage processes the input with a two-dimensional network operating on the STFT representation. Here, we adopt a neural network based on NCSN++, following prior work \cite{richter2023speech, lemercier2025unsupervised}. The inverse STFT is applied at the output to map the intermediate estimate back to the time domain; importantly, this operation is differentiable, enabling end-to-end training.
The second stage is designed to refine early reflections. We extract the early segment of both the input noisy signal and the first-stage time-domain output using a fixed mixing time of $70\,\mathrm{ms}$. These early segments are stacked and refined with a time-convolutional U-Net, which is more capable for modeling sparse, transient structure. 
The final output combines the refined early segment with the late segment produced by the first-stage network. 

To enforce rotation equivariance in Ambisonic RIRs, so results remain consistent regardless of orientation, we apply random rotations and their inverses at the input and output of $F_\psi$. Specifically, we use a Wigner-D rotation matrix $\mathbf{R}_{\phi, \theta}$ where the azimuth $\phi$ and zenith $\theta$ are uniformly sampled over the sphere. During training, this strategy helps reduce overfitting and increases data diversity. 
At inference time, it can further enhance model performance at minimal computational cost, following the \textit{equivariant sampling} strategy \cite{wu2025equivariant}.

\begin{figure}[t]
\resizebox{0.95\columnwidth}{!}{%
\centering
\begin{tikzpicture}[
  font=\scriptsize,
  >=Latex,
  node distance=5mm,
  block/.style={draw, rounded corners=2mm, minimum height=6mm, minimum width=8, align=center},
  Rblock/.style={},
  sum/.style={draw, circle, inner sep=0pt, minimum size=3mm},
  gain/.style={draw, isosceles triangle, isosceles triangle apex angle=60, shape border rotate=-30, minimum height=2.5mm, inner sep=1pt},
  gain_down/.style={draw, isosceles triangle, isosceles triangle apex angle=60, shape border rotate=-120, minimum height=2.5mm, inner sep=1pt},
  split/.style={draw, circle, inner sep=0pt, minimum size=3mm, path picture={
    \draw (path picture bounding box.center) -- ++(-0.1,0)
          (path picture bounding box.center) -- ++(0.07,0.07)
          (path picture bounding box.center) -- ++(0.07,-0.07);
  }},
  concat/.style={draw, circle, inner sep=0pt, minimum size=3mm},
  note/.style={align=center}
]

\draw[rounded corners=2mm, gray, dashed, fill=gray!10] (0.77,-1.4) rectangle (7.73,0.8);
\node[note, gray] at (1.2, 0.55) {\normalsize $F_\psi$};

\node (A0) at (0,0) {$\mathbf{a}_\tau$};

\coordinate (branch) at (0.35,0);

\node[gain] (cin) at (0.55,0) {};
\node[note] at (0.55,0.3) {$c_{\mathrm{in}}$};

\node[Rblock] (R) at (1.35,0) {$\mathbf{R}_{\phi,\theta}$};

\node[block] (stft) at (2.35,0) {$\mathcal{F}_\mathrm{ST}$};
\node[block] (tfnet) at (3.53,0) {NCSN++};
\node[block] (istft) at (4.75,0) {$\mathcal{F}_\mathrm{ST}^{-1}$};

\node[split] (split) at (5.45,0) {};

\node[concat] (merge) at (6.6,0) {$\|$};

\node[Rblock] (Rinv) at (7.30,0) {$\mathbf{R}^{-1}_{\phi,\theta}$};

\node[gain] (cout) at (7.9,0) {};
\node[note] at (7.98,0.35) {$c_{\mathrm{out}}$};

\node[sum] (sum) at (8.45,0) {};
\node at (sum) {$+$};

\coordinate (branch2) at (8.45,0.95);
\node[gain_down] (cskip) at (8.45,0.6) {};
\draw (branch) |- (branch2);
\draw (branch2) -- (cskip.north);
\draw[->] (cskip.south) -- (sum.north);
\node[note] at (8.9,0.5) {$c_{\mathrm{skip}}$};

\node (out) at (9.05,0) {$\hat{\mathbf{a}}_0$};

\draw ($(A0) + (0.2, 0)$) -- (branch);
\draw[-] (branch) -- (cin);
\draw[->] (cin) -- ($(R) + (-0.35,0)$);
\draw[->] ($(R)+(0.3,0)$) -- (stft);
\draw[->] (stft) -- (tfnet);
\draw[->] (tfnet) -- (istft);
\draw (istft.east) -- (split);
\draw[->] (split) -- (merge);
\draw[->] (merge) -- ($(Rinv)+ (-0.3,0)$);
\draw[-] (Rinv) -- (cout);
\draw[->] (cout) -- (sum);
\draw[->] (sum) -- ($(out)+(-0.2,0)$);

\node[note] at (5.8,0.25) {late};

\node[block] (timenet) at (6.1,-0.9) {Time\\U-Net};
\coordinate (timenet_in1) at ([yshift=1.5mm]timenet.west);
\coordinate (timenet_in2) at ([yshift=-1.5mm]timenet.west);
\coordinate (cin_out) at (1.7,0);

\draw[->] (split) |- (timenet_in1);
\draw[->] (cin_out) |- (timenet_in2);
\draw[->] (timenet.east) -| (merge);
\node[note] at (5.10,-0.5) {early};

\end{tikzpicture}
}
\vspace{-5pt}
\caption{Diagram of the model architecture}
\vspace{-14pt}
\label{fig:architecture}

\end{figure}


\vspace{-1pt}
\subsection{Posterior Sampling Inference Algorithm} \label{sec:inference}
\vspace{-1pt}


At inference time, we are given microphone-array measurements $\mathbf{x}$ and the corresponding Array Impulse Response $\mathbf{h}$ (or, equivalently, their frequency-domain representation $\mathbf{H}$).
As introduced in Section~\ref{sec:problem_definition}, we aim to approximate the posterior $p(\mathbf{a}\mid \mathbf{x},\mathbf{h})$. In a score-based formulation, we can sample from this posterior using its score function,
\begin{equation}
\label{eq:posterior_score_decomp}
\hspace{-6pt}\nabla_{\mathbf{a}_\tau}\log p(\mathbf{a}_\tau \!\mid \mathbf{x},\mathbf{h})
=
\nabla_{\mathbf{a}_\tau}\log p(\mathbf{a}_\tau)
\hspace{-1pt}
+
\hspace{-1pt}
\nabla_{\mathbf{a}_\tau}
\hspace{-1pt}
\log p(\mathbf{x}\!\mid 
\hspace{-1pt}
\mathbf{a}_\tau,\mathbf{h}),
\end{equation}
and then integrate the probability flow ODE in \eqref{eq:probabilityflow} to obtain posterior samples.
This decomposition disentangles a device-agnostic diffusion prior (Section~\ref{sec:prior}) from a likelihood term that guides the sampling using the measurements and the known forward operator.

For clarity, we describe our likelihood model in the frequency domain.
Since the DFT along time is an invertible linear change of variables, specifying the likelihood in the time domain, $p(\mathbf{x}\mid \mathbf{a}_\tau,\mathbf{h})$, is equivalent to specifying it in the frequency domain, $p(\mathbf{X}\mid \mathbf{a}_\tau,\mathbf{H})$
, up to a constant Jacobian factor that does not affect inference. 

The score of the log-likelihood $\nabla_{\mathbf{a}_\tau}
\hspace{-1pt}
\log p(\mathbf{X}\!\mid 
\hspace{-1pt}
\mathbf{a}_\tau,\mathbf{H})$
 is intractable in general, and many approximations have been proposed \cite{daras2024survey}. Here we build on Diffusion Posterior Sampling (DPS) \cite{chung2022diffusion} and approximate the likelihood score by differentiating a data-consistency discrepancy evaluated at the current denoised prediction.

Let $\hat{\mathbf{A}}_0(\mathbf{a}_\tau)$ denote the DFT (over time) of the denoised estimate $\hat{\mathbf{a}}_0(\mathbf{a}_\tau)$ at time $\tau$. 
DPS uses the approximation
\vspace{-2pt}
\begin{equation}
\nabla_{\mathbf{a}_\tau}\log p(\mathbf{X}\mid \mathbf{a}_\tau,\mathbf{H})
\approx
-\zeta(\tau)
\nabla_{\mathbf{a}_\tau}
d\left(
\mathbf{X},
\mathbf{H}\hat{\mathbf{A}}_0(\mathbf{a}_\tau)
\right),
\label{eq:dps_measurement}
\vspace{-2pt}
\end{equation}
where $d(\cdot,\cdot)$ is differentiable distance in the measurement domain (with minimum 0 attained when its arguments match),
and $\zeta(\tau)\ge 0$ controls the strength of data consistency along the sampling trajectory. 
The gradient in \eqref{eq:dps_measurement} is taken with respect to the noisy variable $\mathbf{a}_\tau$, and can be computed by automatic differentiation.

We instead enforce consistency in a \textit{range-projected} space.
Concretely, we left-multiply by the (per-frequency) pseudoinverse $\mathbf{H}^\dagger$ and compare range-projected quantities in the Ambisonics domain:
\vspace{-1pt}
\begin{equation}
\hspace{-5pt}
\nabla_{\mathbf{a}_\tau}\!\log p(\mathbf{X}\mid \! \mathbf{a}_\tau,\mathbf{H})
\! \approx \!
-\zeta(\tau)
\nabla_{\mathbf{a}_\tau}
d\!\left(
\mathbf{H}^\dagger\mathbf{X},
\mathbf{H}^\dagger\mathbf{H}\hat{\mathbf{A}}_0(\mathbf{a}_\tau)
\right).
\label{eq:dps_range}
\vspace{-1pt}
\end{equation}
Here, $\mathbf{H}^\dagger\mathbf{X}$ corresponds to the LS estimate, and using \eqref{eq:forward_matrix}, $\mathbf{H}^\dagger\mathbf{H}\hat{\mathbf{A}}_0(\mathbf{a}_\tau)$ is the projection of the current denoised prediction onto the measurement-consistent range subspace.

Inspired by prior work \cite{richter2023speech, lemercier2025unsupervised, thuillier2025hrtf}, we employ a distance based on compressed spectrograms.  
\vspace{-2pt}
\begin{equation}
\hspace{-6pt}
d(\mathbf{A}_\mathrm{ST}, \hat{\mathbf{A}}_\mathrm{ST})
\hspace{-1pt}
=    
\hspace{-1pt}
\frac{1}{(N
\hspace{-2pt}
+
\hspace{-2pt}
1)^2 T_S}  
   \left\lVert
   S_\mathrm{comp}(\mathbf{A}_\mathrm{ST})-
   S_\mathrm{comp}(\hat{\mathbf{A}}_\mathrm{ST})
   \right\rVert_F^2,
\end{equation}
where $T_S$ is the number of STFT  time frames. Let $\mathbf{A}_{\mathrm{ST}} := \mathcal{F}_\mathrm{ST}(\mathcal{F}^{-1}(\mathbf{A})) \in \mathbb{C}^{(N+1)^2 \times T_S \times K_S}$ denote the STFT of the time-domain signal, with $\mathcal{F}^{-1}$ the inverse DFT over time and $\mathcal{F}_\mathrm{ST}$ the STFT. The compressed spectrogram is then defined as $S_\mathrm{comp}(\mathbf{A}_\mathrm{ST}) := |\mathbf{A}_{\mathrm{ST}}|^{\frac{2}{3}} \exp(j\angle \mathbf{A}_{\mathrm{ST}})$.

Magnitude compression makes the distance more sensitive to low-energy features, such as late reverberation and subtle spatial details in RIRs. 
 Using a range-space distance further separates and amplifies the high-order HOA components, which are often weak due to limited microphone coverage. 
As a result, the nonlinear compression more directly emphasizes mismatches in weakly constrained components, improving guidance for subtle spatial structure.

We approximate the posterior score using \eqref{eq:dps_measurement} or \eqref{eq:dps_range} in the decomposition \eqref{eq:posterior_score_decomp}, and integrate the probability flow ODE \eqref{eq:probabilityflow} from $\tau_{\max}$ to $0$ via a first-order Euler solver with stochastic updates \cite{karras2022elucidating}. 
We use equivariant sampling \cite{wu2025equivariant} by applying random rotations at every model evaluation, as described in Section~\ref{sec:architecture}.
We adopt a time-dependent step size $\zeta(\tau)$ 
following \cite{moliner2023solving}.

\vspace{-2pt}
\section{Experiments and Results} \label{sec:experiments} \label{sec:datasets}
\vspace{-5pt}

\vspace{-2pt}
\subsection{Datasets}
\vspace{-1pt}

 All signals are resampled to 24~kHz and are zero-padded to a duration of approximately 1\,s.
We use N3D normalization and ACN channel ordering.
 We experiment with two datasets.

 The first is an internal dataset of wave-based 3D FDTD simulations covering rooms of varying complexity, ranging from simple shoeboxes to intricate geometries with scattering objects, with $T_{60}$ spanning 0.1 to 2.0,s.  The simulations employ a standard rectilinear scheme at the maximum stable Courant number with $f_s = 500\,\text{kHz}$ to ensure negligible numerical dispersion up to $12.5\,\text{kHz}$ over $100\,\text{m}$. The simulations are post-processed using an STFT-based time-varying air absorption filter \cite{kates2020adding}. To generate target signals, we model a virtual spherical cardioid array on a 32nd-order Lebedev grid, downsample outputs to $48\,\text{kHz}$, apply plane wave decomposition via a regularized pseudoinverse \cite{rafaely2015fundamentals} limited to $40\,\text{dB}$ amplification, and truncate the final output to 12th-order Ambisonics.
 We use a room-disjoint 48/5/5 training, validation, and test split with 15 source and 15 receiver positions per room, yielding 1,125 test \acp{rir}.

The second is the publicly available Treble-10 dataset~\cite{mullins2025treble10}, which contains \acp{rir} generated using an 8th-order hybrid simulation (wave-based below 5\,kHz, geometrical above). We adopt a room-disjoint split with 7/1/2 rooms for train/validation/test, and use \textit{Meeting room 1} and \textit{Living room with hallway 1} as the test set.


\vspace{-2pt}
\subsection{Experimental Setup}
\vspace{-1pt}

Diffusion models are trained using the datasets described in Section~\ref{sec:datasets}. Given the relatively small size of these datasets, we apply data augmentation techniques to mitigate overfitting, including small temporal and amplitude jitter, as well as randomized air absorption.
We employ the AdamW optimizer with a learning rate of $1 \times 10^{-4}$, using PyTorch’s default momentum and weight decay parameters. Training is performed with a batch size of 32.
For the FDTD dataset, models are trained for 900k iterations; for Treble-10, 150k iterations. The FDTD model has 102M parameters, the Treble-10 model, 95M.

Evaluation is performed using the Aria Glasses device~\cite{zmolikova2024chime, engel2023project}, which features a 7-microphone array. The \ac{atf} is measured on a KEMAR dummy head. All evaluation experiments are conducted in a controlled simulated environment, where microphone measurements are simulated by applying the \ac{atf} to the Ambisonics \acp{rir} from the test split of each dataset. The original Ambisonics \acp{rir} serve as ground truth for paired evaluation.

 \vspace{-2pt}
 \subsection{Baselines and Ablation Studies}
 \vspace{-1pt}

 We evaluate our method against three baselines. The first is a \textit{Linear Encoder}, specifically the least-squares, time-invariant approach described in Section~\ref{sec:problem_definition}, using diffuse-field equalization (with an order-dependent gain). The second is a time-dependent \textit{Neural Encoder} based on~\cite{heikkinen2024neural}, trained with the Aria Glasses device, using the original architecture and loss, but incorporating residual learning~\cite{deppisch2026residual}. Unlike the original work, our task involves encoding room impulse responses (rather than scenes) and targets higher Ambisonics orders (8th/12th), which are beyond the original scope.
 The third baseline, \textit{Conditional Diffusion}, is a trained to directly approximate $p(\mathbf{a}|\mathbf{x},\mathbf{h})$ for a specific device. Here, the backbone $F_\psi$ is conditioned on $\mathbf{c} = \mathcal{F}^{-1}(\mathbf{H}^\dagger \mathbf{X}) \in \mathbb{R}^{(N+1)^2 \times T}$, which is concatenated to the input $\mathbf{a}_\tau$, and Classifier-Free Guidance~\cite{ho2022classifier} (weight 2) is used to enhance conditioning.

\vspace{-1pt}
\subsection{Objective Metrics}
\vspace{-1pt}

We evaluate the dissimilarity between ground truth and estimated \ac{rir} using two metrics.
The Energy Decay Convergence (EDC)~\cite{dal2024similarity} measures dissimilarity in late reverberation by comparing the energy decay curves of the two signals. For \ac{hoa}, EDC is averaged across all Ambisonics coefficients.
For early reflections (first 100\,ms), we use the Normalized Projection Misalignment (NPM)~\cite{morgan1998onthe}, which quantifies scale-invariant misalignment between the true and estimated signals, also averaged across Ambisonics coefficients.

\vspace{-1pt}
\subsection{Results}
\vspace{-1pt}

The evaluation results for both datasets are summarized in Table \ref{tab:fdtd_treble_side_by_side}.
Across all metrics and datasets, the proposed method demonstrates clear improvements over the linear encoder baseline, achieving higher accuracy in both early and late reverberation characteristics. The neural encoder, in contrast, consistently underperforms, which we attribute to the method is not capable of providing a reliable estimate at higher orders.
The conditional diffusion baseline also yields lower performance compared to our approach. We hypothesize that this is due to the absence of explicit range-space consistency constraints, which are central to the effectiveness of our method. 

In ablation studies, replacing the proposed range-projected distance with a measurement distance in the likelihood objective results in a consistent decrease in performance across all evaluation metrics, underscoring the importance of the range-projection constraint. Furthermore, alternative backbone architectures based solely on STFT or waveform processing exhibit a marked reduction in early reflection accuracy, as indicated by the NPM metric results.


\begin{table}[t]
\centering
\caption{Results on FDTD (left) and Treble-10 (right) (mean $\pm$ std); lower is better. Best in \textbf{bold}, second best in \uline{underline}.}
\vspace{-8pt}
\label{tab:fdtd_treble_side_by_side}
\setlength{\tabcolsep}{4pt}
\renewcommand{\arraystretch}{1.05}
\resizebox{0.9\columnwidth}{!}{%
\begin{tabular}{l|cc|cc}
\toprule
& \multicolumn{2}{c|}{\textbf{Internal FDTD}} & \multicolumn{2}{c}{\textbf{Treble-10}} \\
\textbf{Method} & EDC $\downarrow$ & NPM $\downarrow$  & EDC $\downarrow$ & NPM $\downarrow$  \\
\midrule
Linear Encoder
& 0.049 $\pm$ 0.05 & 0.999 $\pm$ 0.02 
& 0.106 $\pm$ 0.12 & 1.042 $\pm$ 0.04  \\
Neural Encoder
& 0.043 $\pm$ 0.04 & 1.080 $\pm$ 0.01 
& 0.067 $\pm$ 0.01 & 1.120 $\pm$ 0.01  \\ 
Cond. Diffusion
& 0.026 $\pm$ 0.03 & \uline{0.938} $\pm$ 0.07  
& 0.022 $\pm$ 0.03 & 0.912 $\pm$ 0.10 \\ \midrule
Ours
& \textbf{0.018} $\pm$ 0.02 & \textbf{0.874} $\pm$ 0.13 
& \uline{0.019} $\pm$ 0.01 & \textbf{0.803} $\pm$ 0.16  \\
Ours (meas. dist.)
& 0.025 $\pm$ 0.02 & 0.967 $\pm$ 0.07 
& 0.026 $\pm$ 0.02 & \uline{0.849} $\pm$ 0.14  \\
Ours (STFT)
& \uline{0.021} $\pm$ 0.02 & 1.065 $\pm$ 0.01 
& 0.047 $\pm$ 0.02 & 1.083 $\pm$ 0.02  \\
Ours (waveform)
& 0.022 $\pm$ 0.03 & 1.012 $\pm$ 0.04 
& \textbf{0.018} $\pm$ 0.01 & 0.950 $\pm$ 0.11  \\
\bottomrule
\end{tabular}
}
\vspace{-10pt}
\end{table}

\vspace{-1pt}
\subsection{Perceptual Evaluation}
\vspace{-1pt}

 We conducted a listening test to assess the perceptual impact of the proposed method relative to the baselines.
We considered two sources of Ambisonics reference \acp{rir}: three examples from the FDTD test set and three internally measured \acp{rir} from an Eigenmike-64 array. The latter were encoded to 7th-order Ambisonics using regularized least-squares \cite{politis2017comparing}. For each reference \ac{rir}, we simulated microphone array measurements by applying the Aria Glasses \ac{atf}, yielding device-specific signals, and then re-estimated Ambisonics \acp{rir} using the proposed method and the three baselines.

We applied three rotations to the estimated \ac{hoa} \acp{rir}: no rotation ($\phi = 0^\circ$, $\theta = 0^\circ$), azimuth-only ($\phi = 45^\circ$, $\theta = 0^\circ$), and combined azimuth and elevation ($\phi = 90^\circ$, $\theta = 30^\circ$). Applying the rotations after Ambisonics encoding makes the task more challenging, as the binaural \acp{rir} are evaluated for source directions not directly aligned with the original array configuration. All Ambisonics \acp{rir} (reference and re-estimated) were rendered to binaural signals using a generic HRTF set, and a broadband drum-loop signal was convolved with each binaural \ac{rir} to generate the final stimuli.

The listening test followed a multi-stimulus paradigm with an explicit reference. In each trial, listeners were presented with the binaural reference and the corresponding processed stimuli (proposed method and baselines) in randomized order. The reference remained available throughout, and no low-quality anchor was included. Participants rated each processed stimulus on a continuous scale from 0 (very dissimilar to the reference) to 100 (perceptually indistinguishable). Each configuration was repeated twice, conforming 12 trials per experiment. A total of 13 normal-hearing listeners participated.

\begin{figure}
    \centering
    \includegraphics[width=0.9\columnwidth, ]{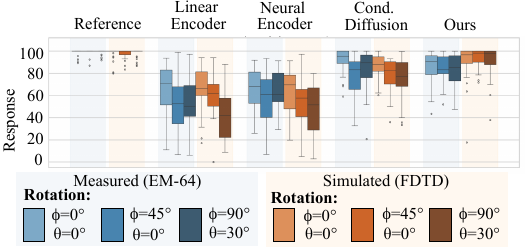}
    \vspace{-5pt}
    \caption{Boxplot representations of the listening test results.}
    \vspace{-15pt}
    \label{fig:test_results}
\end{figure}

Figure~\ref{fig:test_results} shows box plots of the similarity ratings. For the simulated FDTD examples, the proposed method clearly outperformed all baselines, with median ratings close to the top of the scale and minimal degradation under rotation, while the baselines showed a marked drop in quality for rotated conditions. For the measured Eigenmike examples, ratings of the proposed method were overall lower due to the distribution mismatch, but the proposed method and the device-specific conditional diffusion baseline achieved similarly high scores and remained clearly above the linear and neural encoders. These results indicate that the proposed device-agnostic diffusion model generalizes to out-of-distribution measured \acp{rir}, providing more faithful binaural reproduction than other approaches.

\vspace{-2pt}
\section{Conclusions}
\vspace{-2pt}

We have presented a novel method for Ambisonics encoding of \acp{rir} from 
mircophone arrays based on a diffusion-based generative model, trained solely on Ambisonics-domain data. Experiments on simulated data demonstrate that our approach achieves strong performance up to order 12, strongly surpassing the capabilities of a linear encoder.
Furthermore, a subjective listening test with binaural renderings of both simulated and measured \acp{rir} shows that our method achieves high perceptual similarity to reference Ambisonics \acp{rir} while remaining robust under rotated listening conditions.
Crucially, although the evaluation is limited to a single array, the underlying method is device-agnostic. The trained model is independent of the physical array design and can be applied to any configuration for which the \ac{atf} is available, as supported by preliminary internal testing.

Several avenues remain for future research. First, while our current pipeline assumes a highly accurate \ac{atf}, incorporating measurement uncertainty would improve robustness to real-world error.
Secondly, a substantial distribution mismatch remains between the simulated \acp{rir} used for training and real-world measurements, which can present notable performance limitations.
This gap could be addressed either by introducing domain adaptation strategies or by incorporating measured data into training.
Finally, while the computationally intensive iterative sampling is already well-suited for offline applications like data generation, future work could focus on enabling faster inference.





\bibliographystyle{IEEEtran}
\bibliography{strings,refs}

@inproceedings{heikkinen2026beyond,
  author = {M. Heikkinen and A. Politis and K. Drossos and T. Virtanen},
  title = {Beyond omnidirectional: Neural ambisonics encoding for arbitrary microphone directivity patterns using cross-attention},
  booktitle = {Proc. IEEE ICASSP},
  pages = {1--5},
  year = {2026},
  organization = {IEEE}
}

@article{mccormack2022parametric,
  author = {L. McCormack and A. Politis and R. Gonzalez and T. Lokki and V. Pulkki},
  title = {Parametric ambisonic encoding of arbitrary microphone arrays},
  journal = {IEEE/ACM Trans. Audio Speech Lang. Process.},
  volume = {30},
  pages = {2062--2075},
  year = {2022},
  publisher = {IEEE}
}

@inproceedings{thuillier2025hrtf,
  author = {E. Thuillier and J.-M. Lemercier and E. Moliner and T. Gerkmann and V. Välimäki},
  title = {HRTF Estimation using a Score-based Prior},
  booktitle = {Proc. IEEE ICASSP},
  pages = {1--5},
  year = {2025},
  organization = {IEEE}
}

@inproceedings{nawfal2025ambisonics,
  author = {I. Nawfal and S. D. Manias and M. Souden and J. Merimaa and J. Atkins and E. McMullin and S. Pirhosseinloo and D. Phillips},
  title = {Ambisonics Super-Resolution Using A Waveform-Domain Neural Network},
  booktitle = {AES AVAR},
  year = {2025}
}

@inproceedings{chung2022diffusion,
  author = {H. Chung and J. Kim and M. T. Mccann and M. L. Klasky and J. C. Ye},
  title = {Diffusion posterior sampling for general noisy inverse problems},
  booktitle = {ICLR},
  year = {2023}
}

@inproceedings{wabnitz2011upscaling,
  author = {A. Wabnitz and N. Epain and A. McEwan and C. Jin},
  title = {Upscaling ambisonic sound scenes using compressed sensing techniques},
  booktitle = {Proc. IEEE WASPAA},
  pages = {1--4},
  year = {2011},
  organization = {IEEE}
}

@article{karras2022elucidating,
  author = {T. Karras and M. Aittala and T. Aila and S. Laine},
  title = {Elucidating the design space of diffusion-based generative models},
  journal = {Adv. Neural Inf. Process. Syst.},
  volume = {35},
  pages = {26565--26577},
  year = {2022}
}

@misc{milstein2025diffaudiffusionbasedambisonicsupscaling,
  author = {A. Milstein and N. Shlezinger and B. Rafaely},
  title = {DiffAU: Diffusion-Based Ambisonics Upscaling},
  year = {2025},
  eprint = {2510.00180},
  archivePrefix = {arXiv},
  primaryClass = {eess.AS},
  url = {https://arxiv.org/abs/2510.00180}
}

@inproceedings{politis2017comparing,
  author = {A. Politis and H. Gamper},
  title = {Comparing modeled and measurement-based spherical harmonic encoding filters for spherical microphone arrays},
  booktitle = {Proc. IEEE WASPAA},
  pages = {224--228},
  year = {2017},
  organization = {IEEE}
}

@inproceedings{heikkinen2024neural,
  author = {M. Heikkinen and A. Politis and T. Virtanen},
  title = {Neural ambisonics encoding for compact irregular microphone arrays},
  booktitle = {Proc. IEEE ICASSP},
  pages = {701--705},
  year = {2024},
  organization = {IEEE}
}

@article{deppisch2026residual,
  author = {T. Deppisch and Y. Gao and M. Mittal and B. Stahl and C. Hold and D. Alon and Z. Ben-Hur},
  title = {Residual Learning for Neural Ambisonics Encoders},
  journal = {arXiv preprint arXiv:2601.18322},
  year = {2026}
}

@article{lai2025principles,
  author = {C.-H. Lai and Y. Song and D. Kim and Y. Mitsufuji and S. Ermon},
  title = {The principles of diffusion models},
  journal = {arXiv preprint arXiv:2510.21890},
  year = {2025}
}

@article{lipman2022flow,
  author = {Y. Lipman and R. T. Q. Chen and H. Ben-Hamu and M. Nickel and M. Le},
  title = {Flow matching for generative modeling},
  journal = {arXiv preprint arXiv:2210.02747},
  year = {2022}
}

@misc{lee2026solvingroomimpulseresponse,
  author = {K. Y. Lee and N. Meyer-Kahlen and V. Välimäki and S. J. Schlecht},
  title = {Solving Room Impulse Response Inverse Problems Using Flow Matching with Analytic Wiener Denoiser},
  year = {2026},
  eprint = {2602.00652},
  archivePrefix = {arXiv},
  primaryClass = {eess.AS},
  url = {https://arxiv.org/abs/2602.00652}
}

@misc{xu2025arraydpsunsupervisedblindspeech,
  author = {Z. Xu and X. Fan and Z.-Q. Wang and X. Jiang and R. R. Choudhury},
  title = {ArrayDPS: Unsupervised Blind Speech Separation with a Diffusion Prior},
  year = {2025},
  eprint = {2505.05657},
  archivePrefix = {arXiv},
  primaryClass = {eess.AS},
  url = {https://arxiv.org/abs/2505.05657}
}

@article{lemercier2025unsupervised,
  author = {J.-M. Lemercier and E. Moliner and S. Welker and V. Välimäki and T. Gerkmann},
  title = {Unsupervised blind joint dereverberation and room acoustics estimation with diffusion models},
  journal = {IEEE Trans. Audio Speech Lang. Process.},
  year = {2025},
  publisher = {IEEE}
}

@inproceedings{lluis2025blind,
  author = {F. Lluís and N. Meyer-Kahlen},
  title = {Blind Spatial Impulse Response Generation from Separate Room-and Scene-Specific Information},
  booktitle = {Proc. IEEE ICASSP},
  pages = {1--5},
  year = {2025},
  organization = {IEEE}
}

@inproceedings{arellano2025room,
  author = {S. Arellano and C. Yeh and G. Bhattacharya and D. Arteaga},
  title = {Room Impulse Response Generation Conditioned on Acoustic Parameters},
  booktitle = {Proc. IEEE WASPAA},
  pages = {1--5},
  year = {2025}
}

@article{della2025diffusionrir,
  author = {S. Della Torre and M. Pezzoli and F. Antonacci and S. Gannot},
  title = {DiffusionRIR: Room Impulse Response Interpolation using Diffusion Models},
  journal = {arXiv preprint arXiv:2504.20625},
  year = {2025}
}

@article{richter2023speech,
  author = {J. Richter and S. Welker and J.-M. Lemercier and B. Lay and T. Gerkmann},
  title = {Speech enhancement and dereverberation with diffusion-based generative models},
  journal = {IEEE/ACM Trans. Audio Speech Lang. Process.},
  volume = {31},
  pages = {2351--2364},
  year = {2023},
  publisher = {IEEE}
}

@article{mullins2025treble10,
  author = {S. S. Mullins and G. Götz and E. Bezzam and S. Zheng and D. G. Nielsen},
  title = {Treble10: A high-quality dataset for far-field speech recognition, dereverberation, and enhancement},
  journal = {arXiv preprint arXiv:2510.23141},
  year = {2025}
}

@inproceedings{dal2024similarity,
  author = {G. Dal Santo and K. Prawda and S. J. Schlecht and V. Välimäki},
  title = {Similarity metrics for late reverberation},
  booktitle = {Proc. Asilomar Conf. Signals, Syst., Comput.},
  pages = {1409--1413},
  year = {2024},
  organization = {IEEE}
}

@article{morgan1998onthe,
  author = {D. R. Morgan and J. Benesty and M. M. Sondhi},
  title = {On the evaluation of estimated impulse responses},
  journal = {IEEE Signal Process. Lett.},
  volume = {5},
  number = {7},
  pages = {174--176},
  year = {1998}
}

@article{daras2024survey,
  author = {G. Daras and H. Chung and C.-H. Lai and Y. Mitsufuji and J. C. Ye and P. Milanfar and A. G. Dimakis and M. Delbracio},
  title = {A survey on diffusion models for inverse problems},
  journal = {arXiv preprint arXiv:2410.00083},
  year = {2024}
}

@article{wu2025equivariant,
  author = {C. Wu and Q. Kong and P. Zhao and W. Yang and W. Ma and F. Tang and Z. Jiang and S. K. Zhou},
  title = {Equivariant Sampling for Improving Diffusion Model-based Image Restoration},
  journal = {arXiv preprint arXiv:2511.09965},
  year = {2025}
}

@book{zotter2019ambisonics,
  author = {F. Zotter and M. Frank},
  title = {Ambisonics: A practical 3D audio theory for recording, studio production, sound reinforcement, and virtual reality},
  year = {2019},
  publisher = {Springer}
}

@book{rafaely2015fundamentals,
  author = {B. Rafaely},
  title = {Fundamentals of spherical array processing},
  volume = {8},
  publisher = {Springer}
}

@inproceedings{moliner2023solving,
  title={Solving audio inverse problems with a diffusion model},
  author={Moliner, E. and Lehtinen, J. and V{\"a}lim{\"a}ki, V.},
  booktitle={IEEE ICASSP},
  pages={1--5},
  year={2023},
  organization={IEEE}
}

@misc{lin2026genchoroomimpulseresponse,
      title={Gencho: Room Impulse Response Generation from Reverberant Speech and Text via Diffusion Transformers}, 
      author={Lin, J. and Su, J. and Anand, N. and  Jin, Z. and Kim, M. and Smaragdis, P.},
      year={2026},
      eprint={2602.09233},
      archivePrefix={arXiv},
      primaryClass={cs.SD},
      url={https://arxiv.org/abs/2602.09233}, 
}

@inproceedings{ratnarajah2022fast,
  title={FAST-RIR: Fast neural diffuse room impulse response generator},
  author={Ratnarajah, A. and Zhang, S.-X. and Yu, M. and Tang, Z. and Manocha, D. and Yu, D.},
  booktitle={IEEE ICASSP},
  pages={571--575},
  year={2022},
  organization={IEEE}
}

@inproceedings{zmolikova2024chime,
  title={The CHiME-8 MMCSG Challenge: Multi-modal conversations in smart glasses},
  author={Zmolikova, K. and Merello, S. and Kalgaonkar, K. and others},
  booktitle={8th International Workshop on Speech Processing in Everyday Environments (CHiME)},
  pages={7--12},
  year={2024}
}

@article{ho2022classifier,
  title={Classifier-free diffusion guidance},
  author={Ho, Jonathan and Salimans, Tim},
  journal={arXiv preprint arXiv:2207.12598},
  year={2022}
}

@article{engel2023project,
  title={Project aria: A new tool for egocentric multi-modal ai research},
  author={Engel, Jakob and Somasundaram, Kiran and Goesele, Michael and Sun, Albert and Gamino, Alexander and Turner, Andrew and Talattof, Arjang and Yuan, Arnie and Souti, Bilal and Meredith, Brighid and others},
  journal={arXiv preprint arXiv:2308.13561},
  year={2023}
}

@article{hansen1987truncated,
  title={The truncated SVD as a method for regularization},
  author={Hansen, Per Christian},
  journal={BIT Numerical Mathematics},
  volume={27},
  number={4},
  pages={534--553},
  year={1987},
  publisher={Springer}
}

@article{kates2020adding,
  title={Adding air absorption to simulated room acoustic models},
  author={Kates, James M and Brandewie, Eugene J},
  journal={The Journal of the Acoustical Society of America},
  volume={148},
  number={5},
  pages={EL408--EL413},
  year={2020},
  publisher={AIP Publishing}
}

@inproceedings{schusterbauer2025diff2flow,
  title={Diff2flow: Training flow matching models via diffusion model alignment},
  author={Schusterbauer, Johannes and Gui, Ming and Fundel, Frank and Ommer, Bj{\"o}rn},
  booktitle={Proceedings of the Computer Vision and Pattern Recognition Conference},
  pages={28347--28357},
  year={2025}
}

\end{document}